# Gapped nodal planes drive a large topological Nernst effect in a chiral lattice antiferromagnet

Nguyen Duy Khanh[1,2]*, Susumu Minami[1,3], Moritz M. Hirschmann[1], Takuya Nomoto[4], Ming-Chun Jiang[1,5], Rinsuke Yamada[2], Niclas Heinsdorf[6,7], Daiki Yamaguchi[2], Yudai Hayashi[2], Yoshihiro Okamura[1,2], Hikaru Watanabe[4], Guang-Yu Guo[5,8], Youtarou Takahashi[1,2], Shinichiro Seki[1,2,9], Yasujiro Taguchi[1], Yoshinori Tokura[1,2,10], Ryotaro Arita[1,4], and Max Hirschberger[1,2]*

[1] *RIKEN Center for Emergent Matter Science (CEMS), Wako, Saitama 351-0198, Japan*

[2] *Department of Applied Physics and Quantum-Phase Electronics Center, The University of Tokyo, Tokyo 113-8656, Japan*

[3] *Department of Physics, The University of Tokyo 113-8656, Tokyo, Japan*

[4] *Research Center for Advanced Science and Technology, Institute of Industrial Science, The University of Tokyo, Tokyo 153-8904, Japan*

[5] *Department of Physics and Center for Theoretical Physics, National Taiwan University, Taipei 10617, Taiwan*

[6] *Blusson Quantum Matter Institute, University of British Columbia, Vancouver, BC V6T 1Z4, Canada*

[7] *Max Planck Institute for Solid State Research, Heisenbergstrasse 1, 70569 Stuttgart, Germany*

[8] *Physics Division, National Center for Theoretical Sciences, Taipei 10617, Taiwan*

[9] *Institute of Engineering Innovation, The University of Tokyo 113-8656, Tokyo, Japan and*

[10] *Tokyo College, The University of Tokyo, Tokyo 113-8656, Japan*

The electronic structure of compensated antiferromagnets (CAF) has drawn attention for its ability to create large responses, reminiscent of ferromagnets and suitable for data storage and readout, despite (nearly) net-zero spontaneous magnetization. Many of the striking experimental signatures predicted for CAF – such as giant thermoelectric Nernst effects – are enhanced when two or more electronic bands are nearly degenerate in vicinity of the Fermi energy. Here, we use thermoelectric and electric transport experiments to study the electronic structure of the layered, chiral metal $CoNb_3S_6$ in its all-in-all-out CAF ground state and report near-degeneracies of electron bands at the upper and lower boundaries of the first Brillouin zone. Considering non-symmorphic spin-space group symmetries in the non-relativistic approximation for the ordered phase, these near-degeneracies are approximately protected by a lattice translation combined with spin rotation, and are vestiges of nodal planes enforced by a screw axis symmetry in the paramagnetic state. Hot spots of emergent, or fictitious, magnetic fields are formed at the slightly gapped nodal plane, generating the spontaneous Hall and Nernst effects in this CAF. Taking into account more than six hundred Wannier orbitals, our model quantitatively reproduces the observed spontaneous Nernst effect, emphasizes the role of proximate symmetries in the emergent responses of CAF, and demonstrates the promise of ab-initio search for functional responses in a wide class of materials with reconstructed unit cells due to spin or charge order.

**Introduction**

Compensated antiferromagnets (CAF) are a large class of complex materials which have a variety of advantages for (fast) information control [1–3]. The impact of crystal and magnetic symmetries on the electronic band structure of CAF, specifically on their widely spin-split bands and nodal touchings, is a matter of active current research with prospects to enhance their functional thermoelectric and magnetooptical responses [1, 2, 4–10]. While it is known that breaking combined time-reversal and translation symmetry is a precondition for these responses, the importance of proximate symmetries is a more recent focus of interest: for example, spin symmetries that are broken when relativistic spin-orbit coupling is accounted for (spin-space group symmetries) [7, 10–14].

We study the CAF $CoNb_3S_6$, whose twisted magnetic order simultaneously breaks timereversal and inversion symmetry [15–20] (Figure 1a). First, our experimental analysis demonstrates that the large Nernst effect (NE) of $CoNb_3S_6$ cannot be described by spatially uniform emergent magnetic fields, and requires full ab-initio treatment of the electronic structure in a momentum ($k$-) space theory. Second, state-of-the-art numerics in a large magnetic supercell of more than eighty atoms successfully describe the NE of the electron gas, including its temperature dependence. Spin-space group symmetries enforce near-touchings of Fermi surfaces, despite strong exchange splitting of the band structure on the scale of several hundred milli-electron-volts. Third and finally, we discuss the emergence of such hot-spot regions as a cooperative phenomenon of CAF order [15, 17–20] and symmetry-enforced nodal planes in the thermally disordered (paramagnetic) state (Supplementary Note S5). Our combined experimental and theoretical approach demonstrates the promise of modeling the functional responses of a wide array of materials with reconstructed unit cells due to periodic charge or spin orders [21–25].

**Structure, compensated antiferromagnetism, and symmetry**

Figure 1a presents the crystal structure of $CoNb_3S_6$, in which intercalated Co ions impose a $\sqrt{3} \times \sqrt{3}$ super-lattice potential onto the $NbS_2$ sheets. The intercalation of Co ions breaks the parent compound's mirror planes and inversion symmetry, resulting in a hexagonal chiral structure in space group $P6_322$ [16, 26]. In combination with time-reversal symmetry, the screw axis in this space group enforces nodal plane degeneracies at the boundary of the Brillouin zone [27]. Below $T_N = 28$K, this material undergoes a magnetic phase transition to a twisted, all-in-all-out (AIAO) antiferromagnetic structure of magnetic space group $P32'$ [10], as shown in Figure 1c [19, 20]. The magnetic unit cell comprises two layers of intercalated cobalt ions, with eight magnetic sites and a total of eighty atoms. Magnetic spins point approximately inward or outward of a tetrahedron, cancelling to yield nearly zero net magnetization. The magnetic order in the AIAO state breaks the screw symmetry and lifts the two-fold degeneracy at the (former) topological nodal planes (Figure 1b). However, – if we ignore relativistic spin-orbit coupling – there remain three non-symmorphic, spin-only rotations around axes piercing the tetrahedron, so-called spin-space group symmetries of the AIAO state [Fig. 1d], which play an essential role in the discussion of the electronic structure. The generators of AIAO's spin-space group are given in Fig. S11.

The AIAO state can also be described as a superposition of three ordering vectors $\mathbf{Q} = \mathbf{a}^*/2$, $\mathbf{b}^*/2$, and $(-\mathbf{a}^* + \mathbf{b}^*)/2$, where $\mathbf{a}^*$ and $\mathbf{b}^*$ are reciprocal lattice vectors (Figure 4b); this texture realizes a lattice-commensurate and sub-nanometer analog on to larger-scale magnetic skyrmions [19, 20, 28]. Like a skyrmion, the magnetic spins of the AIAO state wrap the unit sphere, distorting the motion of conduction electrons and causing large Hall, Nernst, and magneto-optical effects [29–34].

**Observation of the Topological Nernst effect (TNE)**

Figure 2a, b introduces the spontaneous topological Nernst effect in $CoNb_3S_6$ as a large, thermoelectric, ON/OFF-type transport response originating from CAF domains in absence of sizable net magnetization $M$. The spin texture generates a voltage $V_y$ along the direction mutually perpendicular to the temperature gradient $(-\nabla_x T)$ and the small $M \sim 2 \times 10^{-3} \mu_B$/f.u. of the AIAO order.

The Nernst coefficient $S_{xy} = V_y/(-\nabla_x T)$ at $T = 26$K, just below the transition to the CAF state, reaches $1\mu$V/K – comparable in magnitude to NE in metallic ferromagnets [35–38] – and shows clear hysteresis (±4T) upon sweeping the magnetic field $B$ between ±9T. This hysteresis has been shown to relate to a first-order transition between AIAO and AOAI domains at zero magnetic field, with opposite $B_{em}$ [19]. Consistent with prior work [15, 39], we also find a large spontaneous HE $\rho_{yx} \approx 3\mu\Omega$cm with similar hysteresis below $T_N$, suggesting a common origin for HE and NE [Fig. 2d]. Above the hysteretic regime, $S_{xy}$ and $\rho_{yx}$ gently increase with rising magnetic field, their slopes being shown in Fig. S2. Especially the normal Nernst coefficient $S_0 = dS_{xy}/dB$ is rather small as compared to the spontaneous (zero-field) NE, $6.6 \times 10^{-3} \mu$V/K per Tesla.

The coercive field so rapidly increases upon cooling that, even at 90% of the Neel temperature $T_N$, we are unable to resolve a full hysteresis loop in the range of accessible fields as shown in Figure 2c, d. To address this situation, we carefully prepare a single-domain AIAO (AOAI) state by field-cooling (FC) below $T_N = 28$K in a positive (negative) c-axis field: the corresponding sample history is depicted in Fig. 1e by dashed and solid lines, respectively. At 5K, the field is switched off and transport coefficients can be measured up to above $T_N$ by slowly raising temperature, yielding the spontaneous HE and NE shown in Fig. 2e, f. This large spontaneous Nernst effect in the CAF phase cannot be explained in terms of external magnetic fields or in terms of the net magnetization, unlike in conventional ferromagnetic metals. Instead, the experiments indicate the existence of an internal (effective or *emergent*) magnetic field $B_{em}$ generated by the geometric Berry phase, a property of wave-like propagation of conduction electrons in solids as attributed to the impact of magnetic ordering on the electronic structure. This $B_{em}$ is the physical quantity that bridges the realms of magnetic and electronic structures.

**Origin of the emergent magnetic fields**

Considering $B_{em}$, two limits are discussed in the literature [33, 34, 40]: the real (***r***-) space and the momentum (***k***-) space limits, corresponding the characteristic length scale $\lambda_s$ of spin textures. In a full ***k***-space treatment of electronic structures modified by non-coplanar spin textures, the Hall effect is calculated as a sum $\sim \int d^3\mathbf{k}/(2\pi)^3 B_{em}(\mathbf{k}) f_{FD}(\epsilon_\mathbf{k})$ of occupied states in momentum space, where band crossings and near-degeneracies make large contributions (Fig. 3b) [29, 30, 41]; a similar expression holds for the Nernst effect (Materials and Methods). As full ab-initio modeling of the electronic structure and $B_{em}(\mathbf{k})$ for large magnetic unit cells has remained challenging [21–24, 42–44], approximations have been introduced to model the electronic properties of skyrmion lattices and other complex magnetic structures. Large magnetic textures ($\lambda_s \gg$ lattice unit cell $a$) allow us to define spatially uniform values of the emergent fields $B_{em}$, which are directly proportional to the Hall and Nernst effects according to $\rho_{yx}^T = R_0 B_{em}^{HE}$ and $S_{xy}^T = R_0 B_{em}^{NE}$ [32–34, 43, 45] ($R_0$ and $S_0$ are normal Hall and Nernst coefficients, respectively). In this limit, the effect of $B_{em}$ is opposite in sign for spin-up and spin-down conduction electrons [31, 32, 45] (Fig. 3a).

We demonstrate that – contrary to other canted antiferromagnets or skyrmion lattice hosts – $CoNb_3S_6$ cannot be adequately described by the real-space approximation of spatially uniform $B_{em}$. First, Fig. 3c supports the $\mathbf{k}$-space limit in $CoNb_3S_6$ by comparison to numerical calculations based on a Kondo Hamiltonian [40]. Using reasonable materials parameters (see Materials and Methods), we place $CoNb_3S_6$ in the regime of small skyrmion sizes $\lambda_s$ and long spin-relaxation times $\tau_s$, where $\mathbf{k}$-space theories are most appropriate. Second, Fig. 3d shows good agreement between $B_{em}^{HE}$ and $B_{em}^{NE}$ across a number of materials with spin-chiral magnetic textures, including skyrmion phases, despite their widely disparate materials chemistry (Materials and Methods). $CoNb_3S_6$ is an outlier here, where vastly different $B_{em}$ for NE and HE are obtained when using the real-space approximation; this signifies the breakdown of the real-space approximation.

To further motivate the $\mathbf{k}$-space treatment, we introduce the thermoelectric Nernst and electric Hall conductivities $\alpha_{xy}$ and $\sigma_{xy}$ (Fig. 3e, f, Materials and Methods, Ref. [41]). In ferromagnets, the conductivity ratio $\alpha_{xy}/\sigma_{xy}$ has been established as a sensitive probe for the band-filling dependence of $B_{em}$, with $k_B/e$ (Boltzmann constant, fundamental charge) representing an upper bound allowed by thermal broadening of the Fermi-Dirac distribution [46, 47]. In Fig. 3f, our data approaches the universal value $k_B/e$ at high $T$, confirming the sharp filling dependence of $B_{em}$ on the Fermi energy $E_F$; at low $T$, a linear trend of $\alpha_{xy}/\sigma_{xy}$ is a direct measure for the carrier relaxation time (see Materials and Methods). In summary, we assess that the (thermo-) electric transport of CoNb$_3$S$_6$ should be considered using $k$-space emergent fields, via full ab-initio band theory calculations.

**Theoretical model**

While previous work reported the electronic structure in the paramagnetic state of CoNb$_3$S$_6$ [15, 39, 48–51], including the effect of electron correlations [18], and considered magnetic states other than AIAO [17, 25], there remains a need to perform ab-initio calculations of $\sigma_{xy}^T$ and $\alpha_{xy}^T$ in the AIAO CAF structure, to gain insight into the microscopic origin of topological Hall and Nernst effects. Figure 4a depicts the *lower half* of the magnetic unit cell, as viewed along the *c*-axis: four cobalt, twelve niobium, and twenty-four sulfur ions are crowded into this space, which is four times larger than the crystallographic unit cell (dashed box). Figure 4b illustrates the corresponding shrinkage of the Brillouin zone in *k*-space, as well as three magnetic ordering vectors $Q_i$ ($i$ = 1, 2, 3) generating the AIAO state. First, we have calculated the electron bands of paramagnetic (PM) CoNb$_3$S$_6$, yielding a pair of hole-type, tubular Fermi surfaces around the Brillouin zone center as well as a more three-dimensional, electron-like Fermi surface sheet [Fig. 4c, d; Materials and Methods]. The red shaded areas in Fig. 4d, e mark $k_z = \pm\pi/c$, where nodal degeneracies of the Fermi surface (red lines) are enforced by a combination of time-reversal symmetry and a screw axis in the chiral PM state of CoNb$_3$S$_6$ [27, 52]. Due to significant interplane electron hopping in this intercalated van-der-Waals system, the Fermi surfaces are warped along the $k_z$ direction, with spin-orbit coupling lifting the degeneracy of Kramers pairs, or partnered states under time-reversal, away from $k_z = \pm\pi/c$.

The emergence of long-range AIAO CAF order below $T_N$ breaks both time-reversal and screw symmetries, lifting the degeneracy of Kramers pairs even at $k_z = \pm\pi/c$ [Fig. 4e]. Nevertheless, the splitting of Fermi surfaces is weak, especially around the (now broken) nodal plane at $k_z = \pm\pi/c$; in comparison, the electron bands of the field-aligned ferromagnetic state have much larger band splitting (Supplementary Information).

If spin-orbit coupling were to be completely ignored in the AIAO state, spin and orbital spaces are decoupled. Now, a translation along *a*/2 − *b*/2 (a half unit cell), combined with one of three 180° spin-only rotations in Fig. 1d, is a good symmetry of the spin system. These three nonsymmorphic spin-space rotation symmetries have orthogonal rotation axes and hence can be represented at all $k$ by the anti-commuting Pauli matrices $i\sigma_n$ with $n = x,y,z$. After choosing one of the rotation axes in Fig. 1d as the spin quantization axis, each band is labeled by one of the symmetry eigenvalues $\lambda_\pm = \pm i$ of $i\sigma_z$. It follows that an eigenstate $|E, \lambda_\pm\rangle$ with energy $E$ of the Hamiltonian is related to an orthogonal state $|E, \lambda_\mp\rangle \propto i\sigma_x |E, \lambda_\pm\rangle$. Thus, such a set of spin-space group symmetries, in absence of spin-orbit effects, enforces two-fold band degeneracies for all $k$ [10–14], and – even if we re-introduce coupling of spin and lattice and lift these symmetries – proximate two-fold band degeneracies remain over large sectors of the Brillouin zone, as demonstrated by DFT: Indeed, Fig. 4e shows pairs of cylinders appearing concentrically, where the outermost pair has been nearly pushed beyond the boundary of the Brillouin zone (see Materials and Methods). At all time-reversal invariant momenta (TRIMs) in $k$-space, the tetrahedral spin-space group symmetries remain an exact symmetry in AIAO even in presence of the spin-orbit interaction, perhaps explaining the more closely spaced pinning of cylinders when approaching the center position ($A$-point) of the $k_z = \pi/c$ plane in Fig. 4e.

**Emergent fields and Nernst response**

We have further implemented the local Berry phase method [42, 44] to fully capture zone-folding effects when calculating the emergent magnetic field $B_{em}$ in AIAO (see Materials and Methods). Figure 4f, g presents the filling dependence of $\sigma_{xy}^T$ and $\alpha_{xy}^T$ in AIAO obtained from this ab-initio calculation. We find quantitative agreement between experiment and theory at the Fermi energy $E_F = +15$meV, as highlighted by orange horizontal lines in Fig. 4f, g. In Supplementary Information, we demonstrate good agreement between the predicted and observed temperature dependence of $\sigma_{xy}^T$ and $\alpha_{xy}^T$. These calculations also confirm that shifting $E_F$ has a drastic effect on $\sigma_{xy}^T$, consistent with prior findings using alloyed crystals [16, 39].

To better understand the origin of the transport coefficients in $k$-space, we sum the emergent magnetic field $B_{em}^z(\boldsymbol{k})$ over two-dimensional slabs of the band structure with $k_z =$ const. (Materials and Methods). Figure 4h shows such data, symmetrized with respect to $k_z$, $-k_z$, removing an antisymmetric component of $B_{em}^z(\mathbf{k})$ that is present already in the paramagnetic state (see Materials and Methods). The calculations for $E_F = 0$, +10meV evidence enhanced contributions to $B_{em}$ at $k_z = \pm\pi/c$, where the nodal plane degeneracy is lifted in the AIAO state (orange highlight). We infer that – although there are other contributions – the slight gapping of topological nodal planes with sizable Chern numbers in vicinity of $k_z = \pm\pi/c$ (see Materials and Methods) creates an important contribution to $\sigma_{xy}^T$ and $\alpha_{xy}^T$ in the AIAO phase.

**Discussion**

Previous work has speculated about the appearance of large $B_{em}$ when nodal planes are gapped by symmetry breaking, and toy model calculations for cubic B20 compounds suggested transport signatures dominated by states close to the boundary of the Brillouin zone [27, 52]. Such work mostly focused on the collinear ferromagnetic state, possibly manipulated by a magnetic field, where large exchange splitting may destroy all remnants of nodal plane degeneracies in real materials (Supplementary Information). In contrast, the present work stresses the importance of proximate symmetries, such as the present spin-space group symmetry of AIAO tetrahedra, for generating near-degeneracies in the electronic band structure and large $B_{em}$ when lifting the degeneracy of a (formerly protected) topological nodal plane [10].

For real materials where long-range charge or spin order and mobile electrons coexist, there is a limited number of theoretical attempts to model functional responses in the ***k***-space limit [21– 25, 53]. The present success in describing the topological Hall and Nernst effects of $CoNb_3S_6$ by ab-initio calculations thus motivates further studies on this vast material class. For magnetic metals in particular, the collaborative effect of magnetic order and ***k***-space topology on electron motion, including the role of spin-space group symmetries, warrants further scrutiny.

---

**Materials and Methods**

**Crystal growth.** Powders of elemental Co, Nb and S are mixed together in a silica tube and annealed at 900°C for 5 days. The process is repeated with re-grinding in between, to enhance the quality of the polycrystal [15]. Single crystals of CoNb$_3$S$_6$ are synthesized by chemical vapor transport (CVT) with iodine as a transport agent. CoNb$_3$S$_6$ crystallizes in a hexagonal thin-plate shape, with the largest surface orthogonal to the $c$-direction. The crystal structure is confirmed by powder X-ray diffraction and Rietveld refinement, and the crystal axes are determined by Laue x-ray back-scattering.

**Magnetization and transport measurements.** The DC magnetic response is measured using a superconducting quantum interference device magnetometer (MPSM 3, Quantum Design). For electric transport measurements, samples are cut into rectangular plate-like shape and equipped with electrodes made from silver paste and thin gold wire (ø40$\mu$m). Measurements of longitudinal ($\rho_{xx}$) and Hall ($\rho_{yx}$) resistivities are performed using a Physical Property Measurement System (PPMS, Quantum Design).

**Thermoelectric measurements.** Thermoelectric experiments are carried out using a customized sample stage mounted in a commercial PPMS cryostat (Quantum Design, Inc.). Thermopower (Seebeck) and Nernst effect are recorded using a one-heater two-thermometer technique in steady-state mode. A temperature gradient $-\nabla T$ is applied along the $a$-direction within the basal plane, while the magnetic field **B** is parallel to the $c$-axis [43]. To correct the effect of contact misalignment, $\rho_{xx}$ and $S_{xx}$ ($\rho_{yx}$ and $S_{xy}$) are (anti-)symmetrized against $B$, respectively. We take care to reduce spurious voltages from electromotive forces at metal junctions, and to detect voltages and temperatures for the calculation of $S_{xx}$ at the exactly same position on the crystal. We acknowledge recent thermoelectric measurements, paired with wonderful imaging data [54], on thin-flake devices of CoNb$_3$S$_6$ using the AC technique, where the low-temperature data was reported without field-cooling procedure and the observed $S_{xy}$ is three times smaller than the presently observed values.

**Density functional theory calculations.** The electronic structure of paramagnetic and magnetically ordered CoNb$_3$S$_6$ is calculated using the OpenMX code [S1] - in the paramagnetic case, by the same method as in the ARPES study of Ref. [50]. We use the exchange-correlation functional within the generalized gradient approximation and with norm-conserving pseudopotentials [S2, S3]. The wave functions are expanded by a linear combination of multiple pseudoatomic orbitals [S4, S5]. Spin-orbit coupling (SOC) is included through total angular momentum-dependent pseudopotentials [S6]. A set of pseudoatomic orbital bases was specified as Co6.0-$s_3p_2d_2f_1$, Nb7.0-$s_2p_2d_2f_1$, and S7.0-$s_3p_3d_2$, where the number after each element stands for the radial cutoff in Bohr radii; the integers after $s$, $p$, $d$, and $f$ indicate the radial multiplicity of each angular momentum component. The lattice constants of CoNb$_3$S$_6$ with all-in all-out, non-coplanar magnetic structure are set to $a$ = 11.498 Å and $c$ = 11.886 Å. A charge density cutoff energy of 500Ry and a $\boldsymbol{k}$-point mesh of 6 × 6 × 5 are used. The (small) net magnetization is along the crystallographic $c$-axis.

Electric and thermoelectric conductivities. The electric conductivity and resistivity tensors are defined by the relations $\boldsymbol{J} = \sigma \boldsymbol{E}$ and $\boldsymbol{E} = \rho \boldsymbol{J}$, while the thermoelectric conductivity and Seebeck / Nernst tensors are defined via $\boldsymbol{J} = \alpha(-\nabla T)$ and $\boldsymbol{E} = S(-\nabla T)$, respectively. Here $\boldsymbol{J}$, $\boldsymbol{E}$, and $(-\nabla T)$ are the electric current density, the electric field, and a temperature gradient, respectively. In the formalism of Fig. 3b, the decompositions according to

$$\rho_{yx}^T = R_0 . P . B_{em}^{HE} \tag{1},$$

$$S_{xy}^T = S_0 . P . B_{em}^{NE} \tag{2},$$

with the normal Hall (Nernst) coefficient $R_0$ ($S_0$) and the spin polarization factor $P$ hold under the condition of (a) moderately weak band splitting between ↑ and ↓ states and (b) a spatially uniform $B_{em}$ [33, 34, 45] [S7, S8]. Fig. 3d shows that even the prototypical skyrmion host MnSi violates the proposed scaling at very low temperature, which is attributed to the increasing divergence of normal Hall coefficients $R_0^\uparrow$, $R_0^\downarrow$ for electrons of opposite spin polarization [S7, S8]. In contrast, the CAF Mn$_3$Sn with breathing Kagome lattice has a weakly canted, noncoplanar state for $\boldsymbol{B} \parallel c$, and neatly obeys $B_{em}^{HE} = B_{em}^{NE}$ [S9]. In our simplified geometry, where currents and temperature gradients are two-dimensional vectors confined to a high-symmetry plane of the crystal, $\sigma$ and $\alpha$ are 2 × 2 matrices. For example, $\rho$ = ($\rho_{xx},\rho_{xy};\rho_{yx},\rho_{yy}$) and the off-diagonal part of the conductivity tensor, i.e. the Hall conductivity, is $\sigma_{xy} = \rho_{yx}/(\rho_{xx}^2 + \rho_{yx}^2)$ when setting $\rho_{xx} \approx \rho_{yy}$. Neglecting the off-diagonal thermal conductivity (thermal Hall effect), the thermoelectric conductivity can be related to Seebeck and Nernst effects as $\alpha_{xy} = \sigma_{xx}S_{xy} + \sigma_{xy}S_{xx}$. Mott's relation between thermoelectric and electric conductivity tensors [S10] was recently demonstrated for the topological Nernst effect [34],

$$\frac{\alpha_{ij}}{T} = -\left(\frac{\pi^2}{3}\right)\left(\frac{k_B^2}{e}\right)\left(\frac{\partial \sigma_{ij}}{\partial \epsilon}\right)_{\epsilon=E_F} \quad (3),$$

In consequence, $S_{xy} = -(\pi^2/3)(k_B^2T/e) \times \sigma_{xx}^2/(\sigma_{xx}^2 + \sigma_{xy}^2) \times dtan\theta_H/dE$, where $\theta_H$ is the Hall angle [S11]. According to Mott's relation, the ratio $\alpha_{xy}/\sigma_{xy}$ measures how rapidly $\sigma_{xy}$ changes with band filling [46, 47] – but broadening of the electronic structure by $\Delta E \sim k_BT$ (by disorder $\sim \hbar/\tau$) at low (high) temperature constrains the derivative $\partial\sigma_{xy}/\partial\epsilon$. The limits $(\pi^2/3)(k_B/e)$ and $(\pi^2/3)(k_B^2/e\hbar)\tau T$ for $\alpha_{xy}/\sigma_{xy}$ are obtained by assuming a large change $\Delta\sigma_{xy}/\sigma_{xy} \approx 1$ over these two respective energy intervals.

**Calculation of intrinsic topological Hall and Nernst conductivities.** We use the local Berry phase technique to determine the intrinsic (anomalous or topological) Hall conductivity in the momentum-space limit, avoiding complexities associated with construction of a Wannier representation comprising more than six hundred atomic orbitals [42, 44]. The essential expression is [S12]

$$\sigma_{xy}^{int}(T) = -\frac{e^2}{\hbar}\int_{BZ}\frac{d^3k}{(2\pi)^3}f_{FD}(\mathbf{k},T)B_{em}(\mathbf{k})^z \quad (4),$$

where $f_{FD}(\mathbf{k},T)$ and BZ denote the Fermi-Dirac distribution function and a volume integral over the Brillouin zone, and $B_{em}(\mathbf{k})^z$ is the $z$-component of the emergent magnetic field (Berry curvature) in momentum ($\mathbf{k}$-) space. From this, the thermoelectric Nernst conductivity follows as $\alpha_{xy}^A = (ek_B/\hbar)\int_{BZ}\left[\frac{d^3k}{(2\pi)^3}\right]s(\mathbf{k},T)B_{em}(\mathbf{k})^z$, with $s(\mathbf{k},T)$ the von Neumann entropy density of the electron gas. Here $k_B$, $T$, $e$ and $\varepsilon$ denote the Boltzmann constant, temperature, the fundamental charge, and the band filling (Fermi energy), respectively.

In the PM state with chiral symmetry, combined time-reversal and $C_{2z}$ symmetries ensure opposite sign of $B_{em}^z$ for $k_z$ and $-k_z$. Remnants of this (anti-symmetric) behaviour are visible even in the AIAO calculation (Supplementary Information). To emphasize the main contributors to the nonzero net Hall and Nernst conductivities in Fig. 4f, g, we symmetrize $B_{em}^z(\mathbf{k})$ in Fig. 4h.

**Temperature dependence of calculated Hall and Nernst conductivities.** The intrinsic contribution to the topological Hall conductivity of CoNb$_3$S$_6$ in the low-$T$ limit is calculated by the OpenMX code based on the local Berry phase [44] as in Eq. (4). The intrinsic Hall conductivity $\sigma_{xy}^{T,int}$ and Nernst conductivity $\alpha_{xy}^{T,int}$ at finite temperature are obtained, based on the Boltzmann transport equation and linear response theory, as follows:

$$\sigma_{xy}^{int}(T) = \int d\epsilon \left(-\frac{\partial f_{FD}(T)}{\partial \epsilon}\right) \sigma_{xy}^{int}(\epsilon, T = 0) \qquad (5)$$

$$\alpha_{xy}^{int}(T) = e \int d\epsilon \frac{(\epsilon-\mu)}{T}\left(-\frac{\partial f_{FD}(T)}{\partial \epsilon}\right) \sigma_{xy}^{int}(\epsilon, T = 0) \qquad (6),$$

where $f$ is the Fermi-Dirac distribution function. The numerical integral is performed on a 40 × 40 × 36 **k**-point grid.

**Estimation of coupling strength and mean free path.** In Fig. 3b, CoNb$_3$S$_6$ is placed in the momentum space limit based on comparison of several material parameters: The coupling strength between itinerant and local moments $J \approx 1.0$ eV is estimated from our DFT calculations as the separation energy between spin-up and down states in the partial density of states, consistent with prior work [39]. The mean-free path $l_\text{mfp}$ is calculated as the product of Fermi velocity $v_F$ and relaxation time $\tau$. The former is estimated from angle-resolved photoemission (ARPES), where the linear slope of the band dispersion defines $v_F = \left(\frac{1}{\hbar}\right) \Delta E/\Delta k$ with $v_F = 2.2\times10^5$ m/s for Co-derived bands at the Brillouin zone edge of CoNb$_3$S$_6$ [S13-S15]. The bound for the carrier relaxation time, $\tau > 33$ fs, is obtained from optical conductivity experiments in Fig. S6.

**Calculation of topological charge of nodal plane.** The electronic structure of paramagnetic CoNb$_3$S$_6$ is calculated using the Vienna Ab-initio Simulation Package (VASP) code [S16-S18] based on density functional theory. The generalized gradient approximation of Perdew-BurkeErnzerhof was adopted for the exchange-correlation functional [S19]. In the self-consistent band structure calculations, a plane-wave cutoff of value 500eV and a Γ-centered *k* mesh of 12×12×6 are used. The results are shown in Supplementary Information, section S5 and Fig. S5. Furthermore, the rotation eigenvalues are calculated by the *Irvsp* package [S20], and the results are organized in Tables S1 and S2 for the two nodal planes, respectively.

## ACKNOWLEDGMENTS

We acknowledge fruitful discussions with Kentaro Ueda, Hiroshi Oike, Bruno Kenichi Saika, Taka-hisa Arima. M.M.H. is supported by the Deutsche Forschungsgemeinschaft (DFG, German Research Foundation) under project number 518238332. M.-C.J. and G.-Y.G. are grateful for support from The Ministry of Science and Technology and the National Center for Theoretical Sciences (NCTS) of The R.O.C., as well as RIKEN's IPA Program. We also acknowledge support from the Japan Society for the Promotion of Science (JSPS) under Grant Nos. JP22H04463, JP23H05431, JP23K13058, and 21K13873, as well as from the Murata Science Foundation, Yamada Science Foundation, Hattori Hokokai Foundation, Mazda Foundation, Casio Science Promotion Foundation, Inamori Foundation, and Kenjiro Takayanagi Foundation. This work was partially supported by the Japan Science and Technology Agency via JST CREST Grant Numbers JPMJCR1874, JPMJCR20T1 (Japan), and JST FOREST (JPMJFR2238).

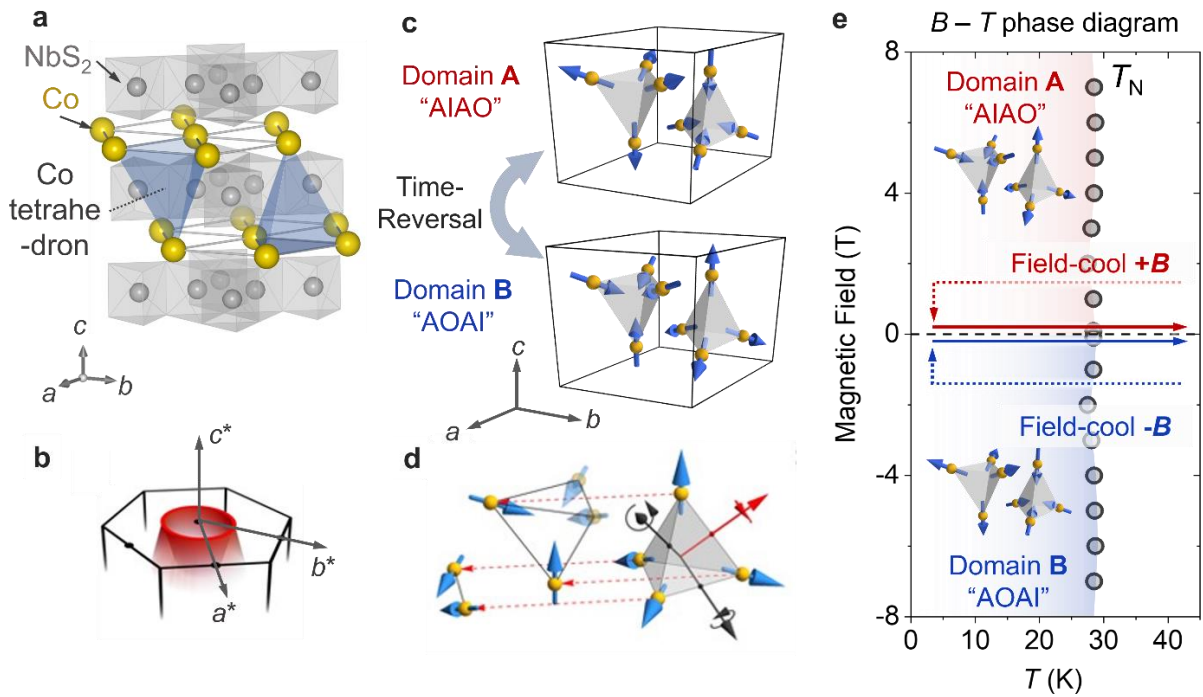

**Figure 1| Structure, magnetism, and symmetry in the compensated antiferromagnet $CoNb_3S_6$. a**, Layered $CoNb_3S_6$ with chiral lattice. **b**, Nodal plane enforced by screw symmetry in the paramagnetic (PM) state (illustration). Black lines, black dots, red surfaces, and red line indicate the boundary of the Brillouin zone, time-reversal invariant momenta at $k_z = \pi/c$, two Kramers-paired Fermi surfaces (FSs), and the intersection of these Fermi surfaces with the nodal plane. **c**, Noncoplanar all-in-all-out (AIAO) magnetic order of cobalt ions (yellow spheres). Two magnetic domains are related by time reversal symmetry [19, 20]. **d**, Proximate, non-symmorphic spin-space group symmetries in the AIAO state: For the three spin-only rotations around red and black axes by 180°, the spin direction rotates while its position (in real space) remains the same. This is followed by spin translation along $a/2 - b/2$, indicated by red dashed arrows corresponding to the red axis. e, Magnetic field ($B$) – temperature ($T$) phase diagram of $CoNb_3S_6$ obtained from magnetization $M(T)$ at various $B$. A single-domain AIAO or AOAI state (inset) can be selected by field-cooling (FC) the sample in positive or negative $B$, respectively. Dashed and solid arrows indicate the sample history during cooling run and measurement run, respectively. $T_N$ is the Neel temperature and the´ magnetic field is parallel to the $c$-axis.

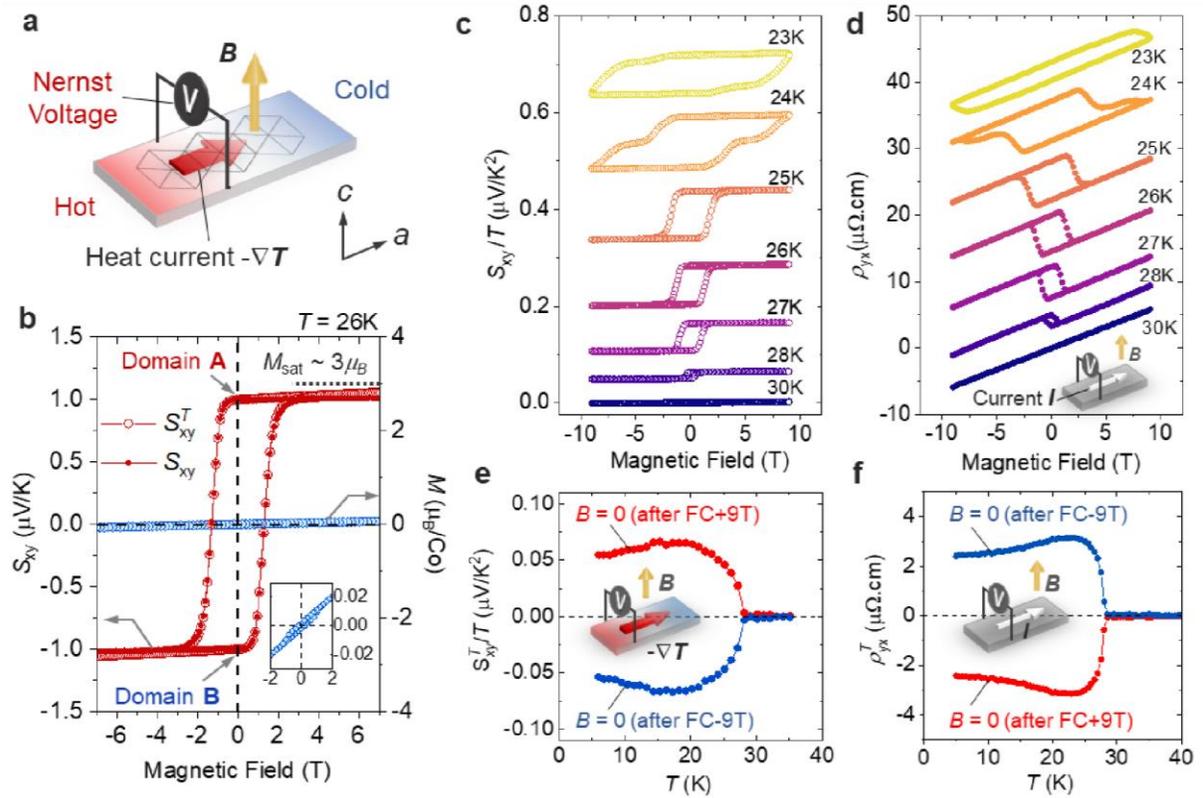

**Figure 2| Large, spontaneous topological Nernst effect (TNE) and topological Hall (THE) effect in $CoNb_3S_6$. a**, Experimental setup (schematic) for measurement of TNE. The magnetic field and temperature gradient are **B** ∥ $c$ and $-\nabla T$ ∥ $a$, respectively. **b**, Isotherms of the total Nernst effect $S_{xy}(B)$ before subtraction of the linear (orbital) part, and topological Nernst effect $S_{xy}^T(B)$ after subtraction (left axis). The right axis shows magnetization $M$ (right axis) at $T = 26K$, just below the Neel temperature $T_N$. Dashed horizontal line: Expected saturation magnetization of $Co^{2+}$ ions. Inset: expanded view of $M$ in the low-$B$ regime, with hysteresis. **c**, **d**, Magnetic field dependence of thermoelectric and electric coefficients around $T_N$, with offset shifts. e, f, $T$-dependence of field-trained THE $\rho_{yx}^T$ and TNE $S_{xy}^T$. Insets: measurement geometries for both.

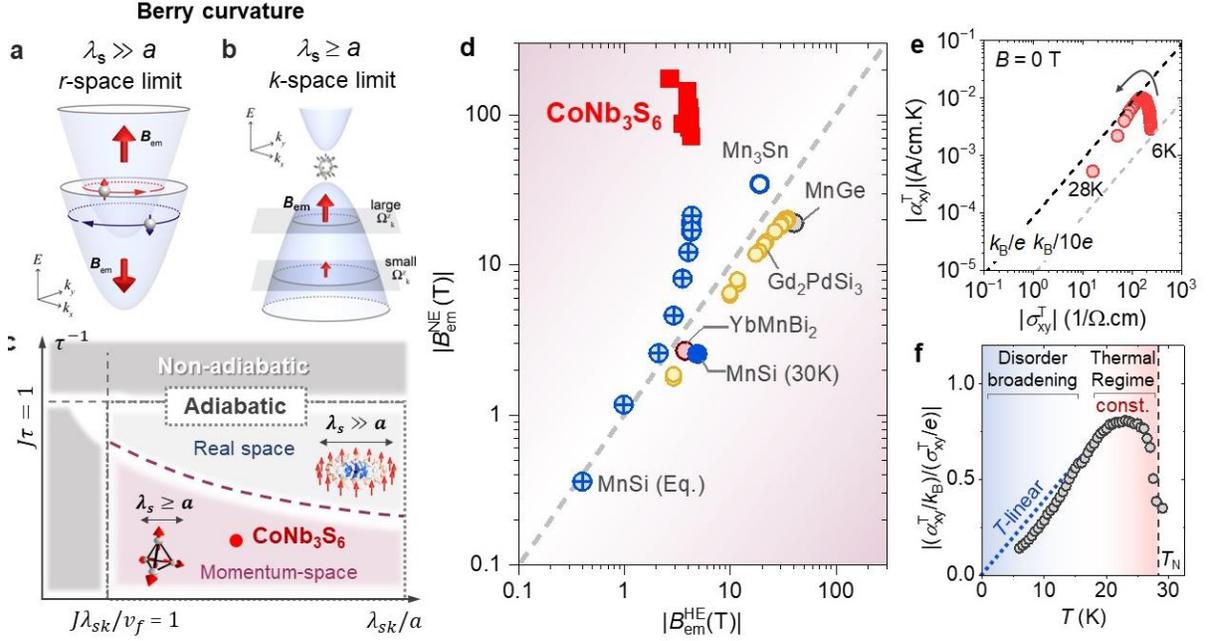

**Figure 3| Emergent electromagnetism driven by spin chirality, and experimental evidence for the momentum-space limit in the CAF CoNb$_3$S$_6$. a**, **b**, Electronic band structure in presence of spatially uniform (real-space) and spatially modulated (*k*-space) emergent magnetic fields $B_{em}$ derived from the Berry phase formalism (qualitative). See text for details. **c**, Placing CoNb$_3$S$_6$ in the parameter space of Matsui *et al.* [40], where $J$, $v_F$, $\tau$, $\lambda_s$, and $a$ are coupling strength between itinerant and local moments, Fermi velocity, carrier relaxation time, and sizes of magnetic and crystallographic unit cells, respectively. **d**, Spatially uniform emergent magnetic fields $B_{em}^{HE}$ and $B_{em}^{NE}$ deduced from topological Hall effect (HE) and Nernst effect (HE), respectively, for various materials with spin textures. MnSi (Eq.) indicates the transport response of the equilibrium skyrmion phase in MnSi. Circles (squares) mark data points where $B_{em}^{HE}$ and $B_{em}^{NE}$ have the same (opposite) sign. CoNb$_3$S$_6$ is an outlier, suggesting the need for full ab-initio calculations to describe its (thermo-) electric transport properties. **e**, **f**, Temperature dependence of topological Nernst and Hall conductivities $\alpha_{xy}^T$ and $\sigma_{xy}^T$ as compared to the universal ratio $k_B/e$ of Boltzmann constant and fundamental charge. Large $\alpha_{xy}/\sigma_{xy}$ around $T_N$ indicates band-filling dependence of the emergent field, while nearly linear-in-$T$ suppression with $T \to 0$ is explained by disorder broadening of the electronic structure (see Methods).

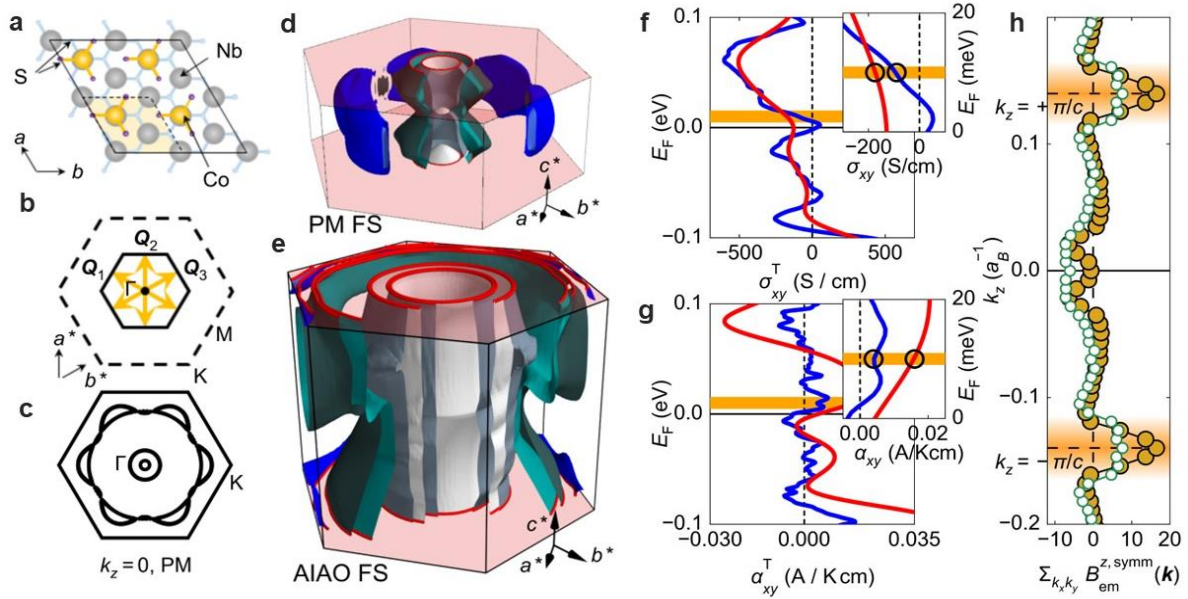

**Figure 4| Electronic structure and modeling of topological Hall and Nernst effects in CoNb$_3$S$_6$. a**, Top view of the unit cell of CoNb$_3$S$_6$, where dashed and solid lines indicate the boundaries of paramagnetic (PM) and all-in-all-out (AIAO) antiferromagnetic unit cells, respectively. **b**, Brillouin zone of PM CoNb$_3$S$_6$ (dashed line) and of the noncoplanar, triple-$Q$ AIAO phase (solid line). **c, d**, Cut of the PM Brillouin zone at $k_z = 0$ plane and its full, three-dimensional view. **e**, Fermi surface and first Brillouin zone in the magnetically ordered state. In **c-e**, the Fermi energy is $E_F = 0$ meV and red shading indicates nodal planes at $k_z = \pm\pi/c$, protected by chiral $P6_322$ symmetry in PM, and slightly gapped in AIAO. Red lines are cross-sections of Fermi surface sheets with $k_z = \pm \pi/c$. **f, g**, Calculated Hall and thermoelectric Nernst conductivities $\sigma_{xy}^T$ and $\alpha_{xy}^T$ as function of $E_F$. Red and blue lines show calculation results at high and low temperature; changes of the ordered moment with $T$ are discussed in Fig. S8. Orange shading indicates the regime at $E_F = +15$ meV with good agreement between theory and experiment. Insets: expanded view at low $E_F$ values. **h**, Sum of z-component of emergent magnetic field $B_{em}^z(\mathbf{k})$ over the $k_x - k_y$ plane, for AIAO. Two values of the Fermi energy $E_F = 0, 10$ meV are labeled in yellow and green, respectively. Anomalies due to gapped nodal planes at $k_z = \pm\pi/c$ are highlighted in orange. The paramagnetic (AIAO ordered) band structures are both calculated with spin-orbit coupling.